\documentclass[aps,pra,twocolumn,preprintnumbers,showpacs,superscriptaddress,amsmath,amssymb]{revtex4}
\usepackage{dcolumn}
\usepackage{bm}
\usepackage{amssymb}
\usepackage[german, english]{babel}
\usepackage{graphicx}
\usepackage{color}
\usepackage{textcomp}
\usepackage[letterpaper,total={7in,9.5in},top=0.75in,left=0.75in]{geometry}
\usepackage[utf8]{inputenc}

\usepackage{array}

\begin{document}

\title{Frequency Conversion in High-Pressure Hydrogen}

\author{Alireza Aghababaei}
\affiliation{Physikalisches Institut, Rheinische Friedrich-Wilhelms-Universität, 53115 Bonn, Germany}
\author{Christoph Biesek}
\affiliation{Physikalisches Institut, Rheinische Friedrich-Wilhelms-Universität, 53115 Bonn, Germany}
\author{Frank Vewinger}
\affiliation{Institut für Angewandte Physik, Rheinische Friedrich-Wilhelms-Universität Bonn, Bonn, Germany}
\author{Simon Stellmer}
\email{stellmer@uni-bonn.de}
\affiliation{Physikalisches Institut, Rheinische Friedrich-Wilhelms-Universität, 53115 Bonn, Germany}

\date{\today}

\begin{abstract}

State-preserving frequency conversion in the optical domain is a necessary component in many configurations of quantum information processing and communication. Thus far, nonlinear crystals are used for this purpose. Here, we report on a new approach based on coherent anti-Stokes Raman scattering (CARS) in a dense molecular hydrogen gas. This four-wave mixing process sidesteps the limitations imposed by crystal properties, it is intrinsically broadband and does not generate an undesired background. We demonstrate this method by converting photons from 434 nm to 370 nm and show that their polarization is preserved.

\end{abstract}

\maketitle

\section{Introduction}
\label{sec:Introduction}
Many schemes for quantum computing and communication are based on a hybrid or distributed architecture and require state-preserving frequency conversion of single photons in the optical domain \cite{Rutz2017,Zaske2012,Krutyanskiy2017,Fisher2021,Bell2016,Ma2012,Tyumenev2022}. Thus far, crystals with a $\chi^{(2)}$ nonlinearity are used for this purpose \cite{Rutz2017,Zaske2012,Ma2012,Fisher2021,Bell2016}. After decades of development, near-unity conversion efficiencies, entanglement-preserving conversion, and a drastic reduction in background have been achieved.

Still, frequency conversion in nonlinear crystals is plagued by various limitations, most prominently the very small spectral acceptance bandwidth of only a few 10 GHz, the generation of undesired optical background, and the intrinsic property of the crystal to convert only one polarization. In addition, there are various technical difficulties, including absorption losses and power handling in the crystal bulk. Crystal properties, such as absorption and birefringence, severely constrain the spectral region in which frequency conversion can be performed.

Here, we explore a pathway to overcome all of these limitations and difficulties. Specifically, we propose to employ a four-wave mixing process in dense molecular gases for frequency conversion. This approach brings the well-established technology of coherent anti-Stokes Raman scattering (CARS) \cite{Zheng2015} to the realms of single photons in quantum information. This process is intrinsically broadband and polarization-insensitive, it does not require the fabrication of delicate waveguide structures and is free of absorption losses, degradation, or power limitations.

The four-wave mixing process is resonantly enhanced in the vicinity of a vibrational transition in a molecule. To this end, the difference frequency of two co-propagating pump fields is adjusted to match a resonance of the molecular gas. The wide selection of molecules, isotopic composition, and transitions allows to bridge a broad range of frequency differences in the range of a few 10 to 100 THz.

In this letter, we demonstrate this approach on a very specific combination of wavelengths, namely 434 nm and 370 nm. Flourene donors in ZnSe can be used as a wavelength-tunable, deterministic source of correlated photons pairs at 434 nm \cite{Sanaka2012, Sleiter2013}. Ytterbium ions, on the other hand, are a well-established qubit for gate operations and have their strongest transition at 370 nm \cite{Kobel2021}. Interfacing these solid-state and atomic platforms is a promising pathway towards a hybrid quantum network. Conversion between these two wavelengths is achieved near the $Q_1$ branch in molecular hydrogen. This conversion would be very challenging in a conventional crystal-based sum-frequency generation process due to the large mismatch between the pump field at 2.4 µm and the signal photons in the ultraviolet. The efficiency of this proof-of-principle experiment is far from being optimized, and does not yet operate with single photons, but with faint laser light.

While preparing this manuscript, we became aware of related work at the Max-Planck Institute for the Science of Light \cite{Tyumenev2022}. There, R.~Tyumenev and co-workers showed state-preserving frequency conversion of infrared photons in a hydrogen gas at a carefully chosen pressure to optimize that phase matching. Confining the modes of the light fields in a photonic crystal fiber allowed the researchers to reach a conversion efficiency of 70 \%. By contrast, our experiment operates in the ultraviolet part of the spectrum.

\section{Experimental setup}
\label{sec:setup}

\begin{figure*}[ht!]
\centering
\includegraphics[width=\textwidth]{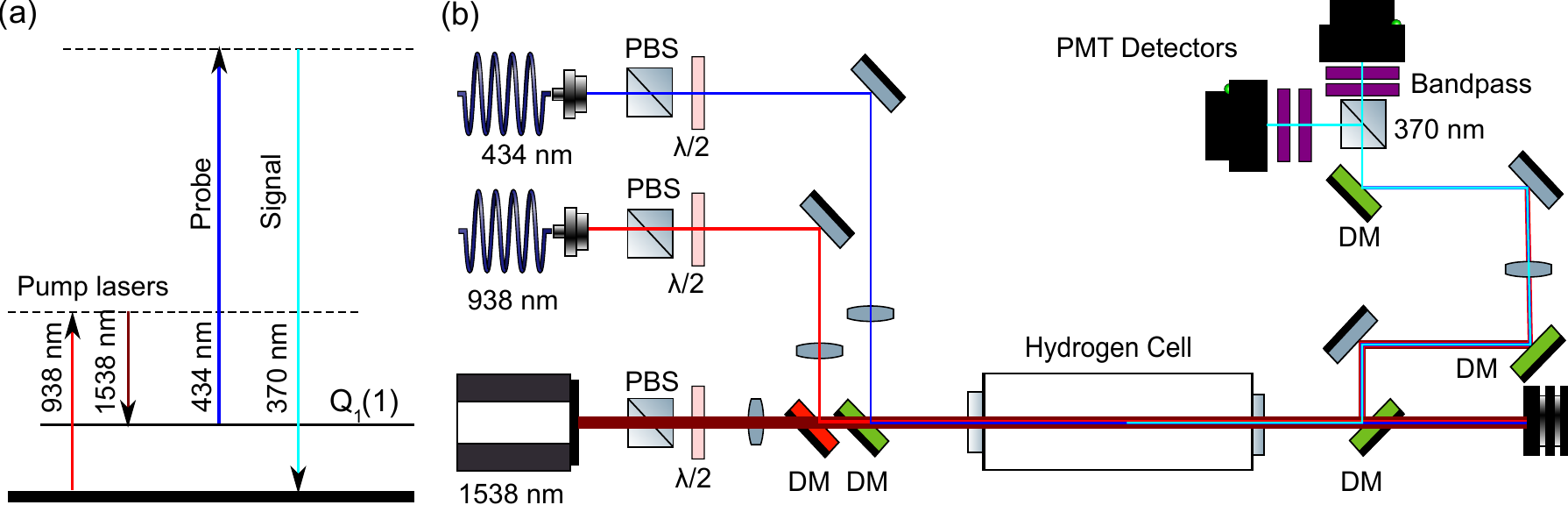}
\caption{Scheme of the frequency conversion process. (a) Energy diagram showing the four light fields involved. The two pump fields are matched in frequency to the $Q_1(1)$ transition in the hydrogen molecule near 124 THz. (b) Experimental setup, showing the pump fields at 1538 nm and 938 nm being overlapped with the signal input at 434 nm via dichroic mirrors (DM) and focused into the hydrogen cell. Two photomultiplier tubes (PMTs) allow for polarization-sensitive detection of the converted photons at 370 nm.}
\label{fig:setup}
\end{figure*}

The heart of our experiment is a home-built, cylindrical high-pressure vessel made from stainless steel, which can hold the hydrogen gas up to a pressure of 60 bar. The cell has a bore with a diameter of 12 mm and length of 140 mm, and is capped by two uncoated sapphire windows. Two gauges, covering different ranges, allow to read off the pressure.

The choice of the pump wavelengths was guided by the availability of high-power continuous-wave laser sources. The first pump laser is an amplified diode laser operating at 938 nm, where about 0.5 W are available to the experiment. The second pump laser (traditionally called the Stokes laser) is a diode laser at 1538 nm whose power is amplified to 16 W by a fiber amplifier. Both lasers are referenced to a commercially available wavelength meter and stabilized in frequency at the MHz-level. Their frequency difference is tuned to match the 124-THz splitting of the $Q_1$ branch \cite{Buijs1971}; see Fig.~\ref{fig:setup}(a). The two pump light fields are focused to a typical waist 70 µm within the hydrogen cell through individual lenses, and are overlapped on a dichroic mirror; see Fig.~\ref{fig:setup}(b) for a schematic representation of the setup.

The probe photons are provided by a laser at 434 nm. A beam with a typical power of a few mW is overlapped with the pump lights and focused to a similar waist. The polarisation of all three beams can be set individually, and their powers are constantly monitored.

Behind the hydrogen cell, the frequency-converted signal photons at 370 nm are separated from the three light fields through a set of three consecutive dichroic mirrors. Further spectral purification is achieved through a series of three bandpass filters, centered at 370 nm with a width of 6 nm and a transmission of 90\% each. The photons pass through a polarizing beam-splitter (PBS) and are detected by two photomultiplier tubes (PMTs, Hamamatsu model H10682-210) with a quantum efficiency of 27 \% and dark count rates of a few counts per second (cps). For alignment and characterization purposes, an auxiliary laser at 370 nm can be overlapped with the 434 nm light.

\section{Frequency conversion}
\label{sec:results}

\subsection{Resonance}

\begin{figure}[b!]
\centering
\includegraphics[width=\columnwidth]{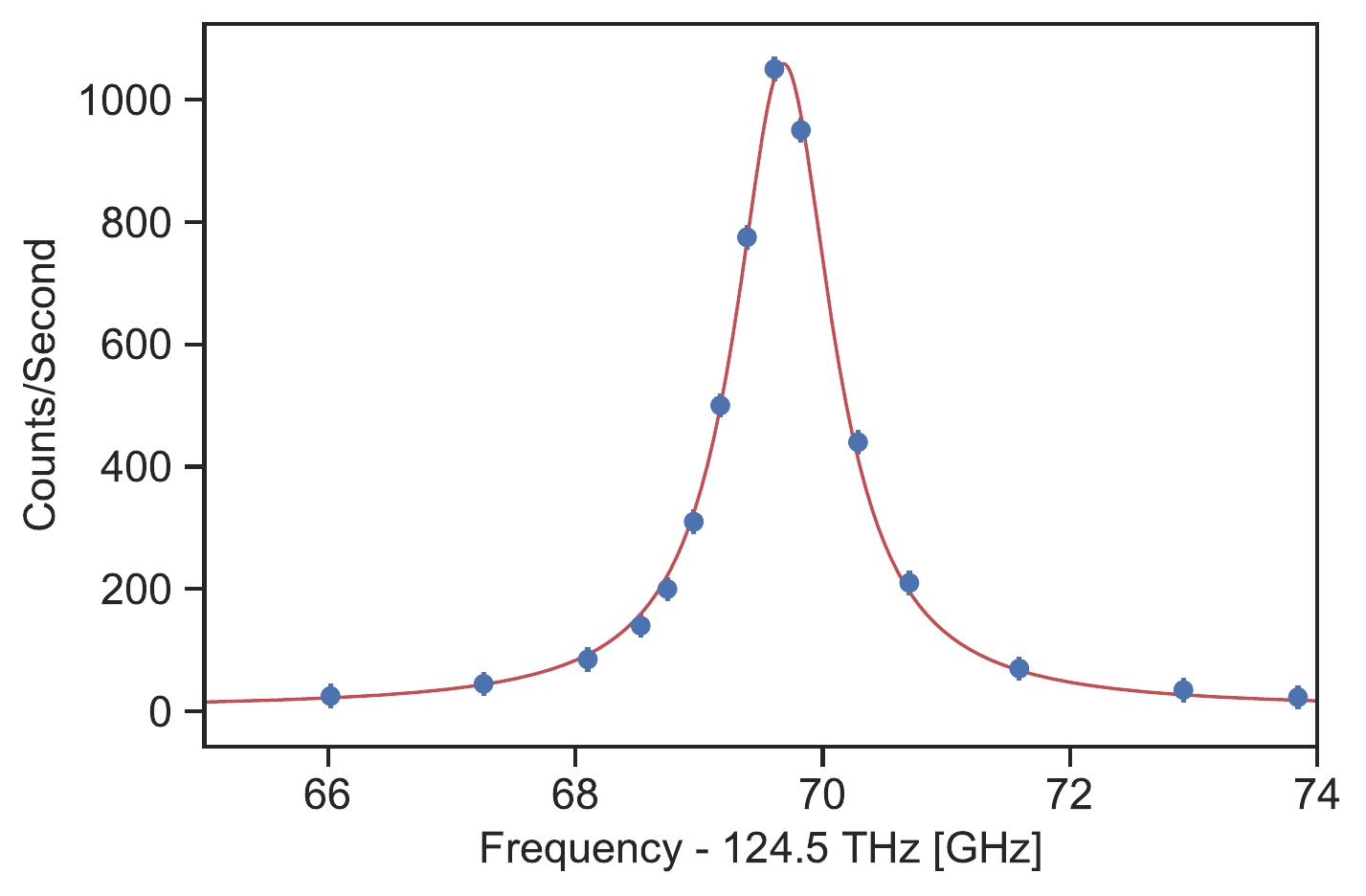}
\caption{Efficiency of the frequency conversion in dependence of the detuning between the two pump fields: a clear resonance at the $Q_1$ transition near 124 THz is observed. The data, taken at a hydrogen pressure of 16.7 bar, is well described by a Lorentzian lineshape. Here, $\nu$ = 124.569683 (5) THz.}
\label{fig:resonance}
\end{figure}

We adjust all three light fields to have the same polarization and similar waists of 70 µm at their focii. At a pressure of 16.7 bar, we scan the relative detuning between the two pump fields and record the rate of frequency-converted photons at 370 nm. A clear resonance is observed; see Fig.~\ref{fig:resonance}.
The position and width of the resonance are pressure-dependent. For the resonance position, we derive a pressure (collisional) shift of $d\nu / dP = -80.0 (1)$ MHz/bar, which is slightly lower than earlier reported values of the shift of the $Q_1(1)$ resonance \cite{Bischel1986,Looi1978}. Extrapolating the data to zero pressure, we find the unperturbed resonance at $\nu_{Q_1} = 124.571055 (5)$ THz, in good agreement with literature values. 

The width of the resonance is a convolution of the natural linewidth and pressure (collisional) broadening. For the comparably high pressures used here, the width of the Lorentzian profile is dominated by linear pressure broadening, with a FWHM-dependence of $40 (1)$  MHz/bar, in agreement with earlier work \cite{Bischel1986,May1961}.

In the experiment, fluctuations and drift in pressure, $dP$, shall be so small that the resonance frequency does not change by more than a fraction of the resonance width, $d\nu / dP \ll \sigma(P)/P$. From the values above, we obtain $dP/P \ll 0.4$. The pressure needs to be stable at the level of a few per cent, which is easily achieved in the experiment.

\subsection{Efficiency}

\begin{figure}[bt!]
\centering
\includegraphics[width=\columnwidth]{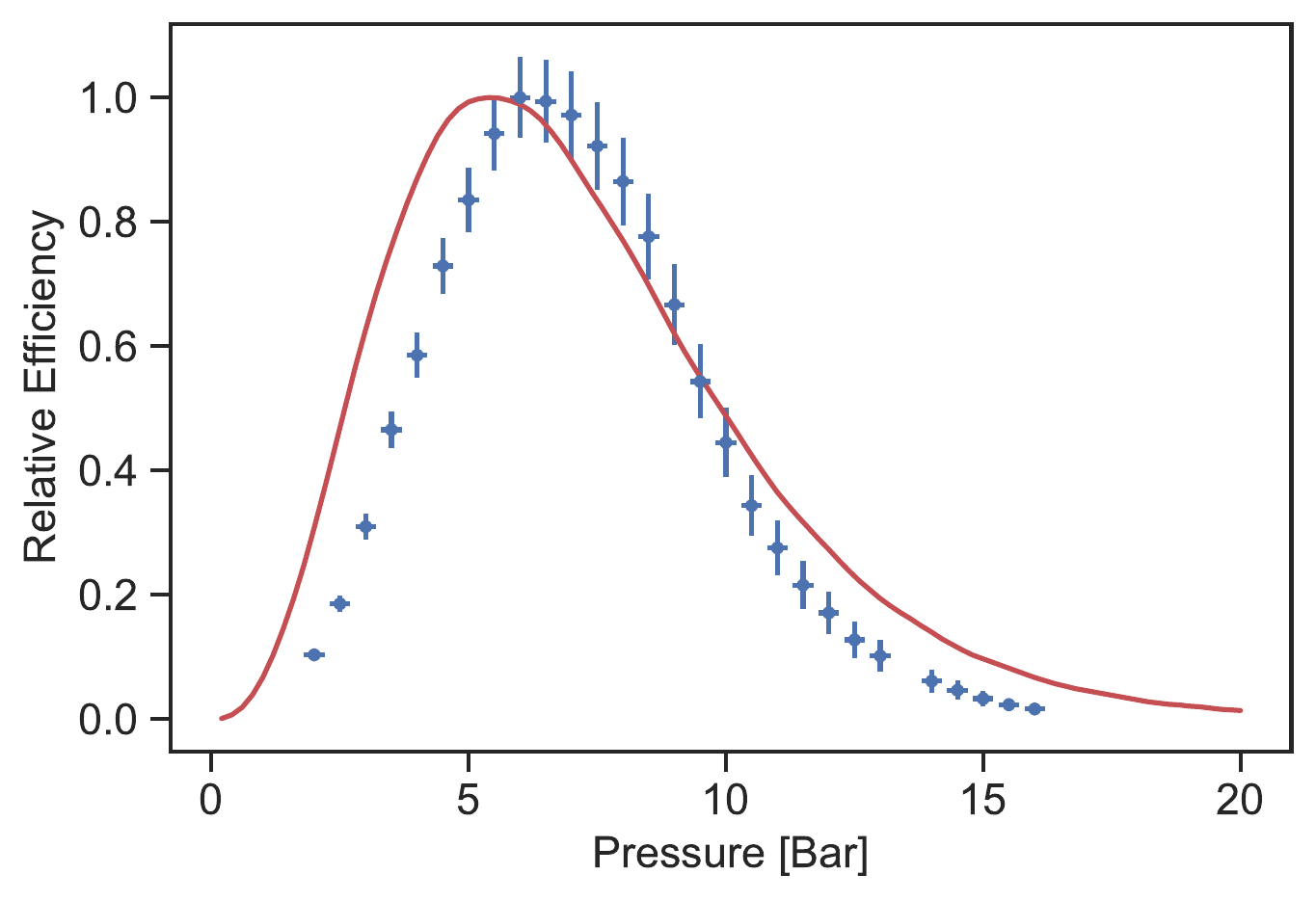}
\caption{Dependence of conversion efficiency on the pressure of the hydrogen gas, which determines the phase mismatch between the light fields, the data are scaled to the maximum. The solid line is the expectation for stimulated Raman scattering in the linear regime.}
\label{fig:pressure}
\end{figure}

The measured efficiency of the conversion process for pressures between 0 and 16 bar is shown in Fig.~\ref{fig:pressure}, together with the expected efficiency for a beam waist of 50 $\mu$m for all  beams. A clear maximum around 6.5(2)  bar is observed, with a FWHM width of approximately 6 bar. 
At a pressure of 6.5 bar, common beam focii of 70 µm, $P_1$ = 15 W and $P_2$ = 500 mW, we obtain an internal conversion efficiency of $1.0 (2) \times 10^{-9}$. Transmission losses on the sapphire windows, dichroic mirrors, and filters lead to a reduced external efficiency of $5.7 (1.1) \times 10^{-10}$. The quantum efficiency of our detectors is about 27\%: the overall probability of in incoming 434-nm photon to be converted and detected is $1.4 (3) \times 10^{-10}$. We experimentally find that the conversion efficiency depends linearly on the power in each of the pump fields, indicating that we are still in the linear regime for stimulated Raman scattering, and correspondingly an increase in pump power can increase the efficiency substantially.

To estimate the relative efficiency, we calculate the electric field of the signal by integrating 
$$\frac{\partial E_{370}}{\partial z} \propto \rho^2 \rho E_{P_1}(r,z) E_{P_2}(r,z) E_\text{Probe}^*(r,z) e^{i \Delta k z},$$ assuming Gaussian beam profiles for the pump and the probe fields \cite{Tolles1977}. Here, $$\Delta k(\rho)=k_{P_1}(\rho)-k_{P_2}(\rho)+k_{\textrm{Probe}}(\rho)-k_{\textrm{Signal}}(\rho)$$ is the phase mismatch that depends on the density $\rho$ of H$_2$ in the cell, and $E_i$ denotes the electric field of the pump and probe fields, respectively. We find the intensity of the signal field by numerical integration. The refractive index at standard conditions is taken from \cite{Peck1977}, and the pressure dependence of $n(\lambda,P)$ is determined using the Lorentz-Lorenz relation, where the number density is determined using the ideal gas law \cite{Diller1968}.

We also mildly vary the focii of the three input light fields, but do not find a clear optimum. For the probe beam this is expected, as its Rayleigh range is much longer than that of the two pump beams.From the simulations one finds that for the pump fields, an decrease in beam waist shifts the maximum to higher pressures, and also leads to a broadening of the peak. This is accompanied by a strong nonlinear increase in  overall efficiency. For the probe beam, a change in beam waist does not significantly alter the efficiency, as its Rayleigh range is much larger compared to those of the two pump fields.

\subsection{Preservation of polarization}

In quantum frequency conversion processes, the polarization of the photon must be preserved. In nonlinear crystals, this requirement cannot easily be fulfilled, as the crystal structure breaks the symmetry between orthogonal polarizations. In our case, the hydrogen gas is isotropic, and the reference frame is set by the polarization of the pump light fields. Adjusting the linear polarization of the pump fields to 45° allows us to convert horizontally and vertically polarized input photons with equal probability.

\begin{figure}[t!]
\centering
\includegraphics[width=\columnwidth]{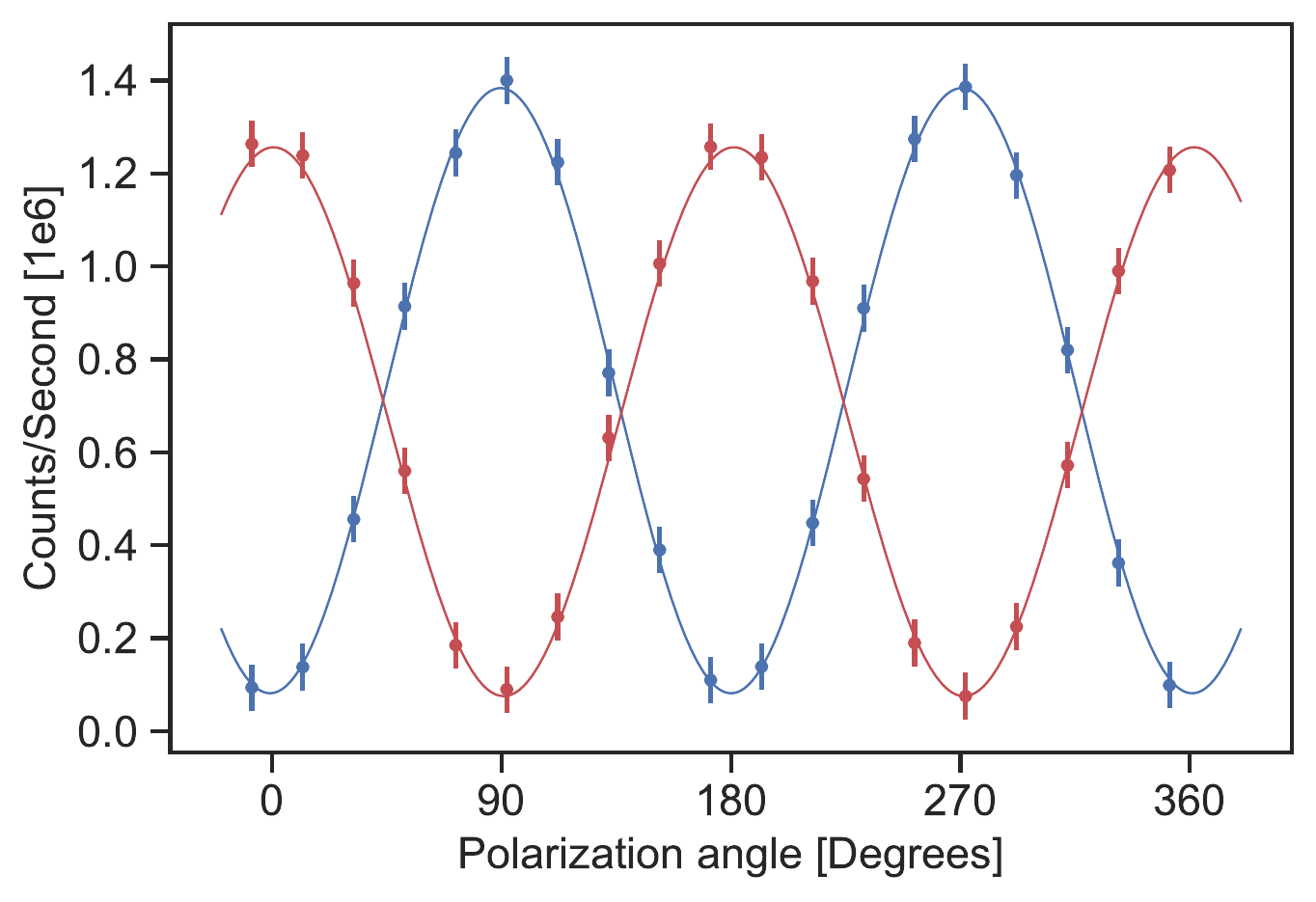}
\caption{A photon retains its polarization through the frequency conversion process. As the polarization of the input light at 434 nm is rotated, the signal in two 370-nm detection channels at orthogonal polarization (red and blue color) follows accordingly with an average fidelity of 93.7(1.6) per cent. Two sine functions are fit to the data.}
\label{double_polarization}
\end{figure}

With the pump fields set to 45°, we vary the polarization of the 434-nm probe photons and monitor the polarization of the frequency-converted signal photons at 370 nm through a polarizing beamsplitter (PBS) and detectors in the two arms. As shown in Fig.~\ref{double_polarization}, polarization is preserved with a fidelity of 93.7(1.6) per cent. The small deviation from unity is entirely explained by imperfect preparation of the input polarization at 434 nm, and the finite 20:1 extinction ratio of the PBS at 370 nm. The difference in amplitude for the two polarizations is caused by slightly different quantum efficiencies of the PMTs. This faithful preservation of polarization fulfills an important prerequisite for the frequency conversion of a quantum state encoded in the polarization state.

\subsection{Background}

Frequency conversion in nonlinear crystals is often empoisoned by the generation of undesired, incoherent background radiation in the wavelength region of interest. Great care needs to be taken to reduce this background count rate, including an optimized choice of crystal material, pumping power, and ultranarrow spectral filtering. Oftentimes, a compromise in conversion efficiency needs to be made in order to optimize signal-to-noise. Here, we show that the conversion process in molecular hydrogen is effectively free of such background processes.

The detectors have average dark count rates of 3.5 cps, which drift by about 3 per cent on intermediate timescales of minutes to hours, likely due to temperature fluctuations, electronic noise, and residual stray light illumination. The detectors' photocathodes are largely insensitive to the IR radiation of the pump fields. A set of dichroic mirrors and bandpass filters allow to suppress any spurious signal of the pump fields to well below 0.1 cps. Now, we aim to detect whether the two pump fields \emph{together} create any background radiation in the UV. We toggle the pump lights on and off to search for an increase in counts rate, but the effect, if any, is below the 0.1 cps random fluctuations and mid-term drifts of the detector dark counts. This is a very promising finding.

\subsection{Downconversion process}

Thus far, we have investigated only the ``upconversion'' process $\nu_{\textrm{Signal}}=\nu_{\textrm{Probe}}+\nu_{P_1}-\nu_{P_2}$, which is relevant for quantum links between ZnSe donors and Yb$^+$ ions and converts photons from 434 nm to 370 nm. The corresponding ``upconversion'' process $\nu'_{\textrm{Signal}}=\nu_{\textrm{Probe}}-\nu_{P_1}+\nu_{P_2}$ is equally possible and generates photons at 530 nm from a 434-nm source.
To study this interaction, we slighly modify the experimental setup downstream of the hydrogen cell by exchanging dichoic mirrors and filters. As before, single 530-nm photons are detected by the PMT, albeit at a reduced quantum efficiency.

We find an internal conversion efficiency of $2.5 (5) \times 10^{-9}$, a 2.5-fold increase compared to the ``upconversion'' process that is fully explained by the improved phase matching at this combination of wavelengths. Similarly, the optimum hydrogen pressure is shifted to a slightly higher value, 7.4 bar.

\section{Conclusion}
\label{sec:conclusion}

In this work, we have proposed and demonstrated a fruitful link between laser technology and quantum information processing. Specifically, we have employed a coherent four-wave mixing process in a dense gas of molecular hydrogen to convert photons from 434 nm to 370 nm. These wavelengths are relevant for hybrid architectures that couple flourene emitters in ZnSe to Yb$^+$ ions. The efficiency, currently at the level of $10^{-9}$, can be improved dramatically by proper choice of the pump wavelengths to minimize the phase mismatch. As shown in recent related work, a near-unit efficiency can be reached when the pump wavelengths are chosen to be similar to the probe and signal wavelengths \cite{Tyumenev2022}. Further increase in conversion efficiency can be achieved by an increase in pump laser power, e.g., through resonant build-up cavities. With these adaptations, conversion and detection of single UV photons will be feasible.

We show that this interaction preserves polarization and thus fulfills an important prerequisite for application in quantum frequency conversion. At our level of sensitivity, the conversion process does not generate an incoherent background.

This approach opens up new avenues of frequency conversion in parameter ranges not accessible by nonlinear crystals, e.g., due to absorption and adverse phase-matching conditions. Extending this approach to the hollow-core fiber platform \cite{Tyumenev2022} will allow for an integrated, highly stable, and very efficient device.

\begin{acknowledgments}
We thank all members of the ML4Q Cluster of Excellence for stimulating discussions, especially Michael Köhl, Stefan Linden, and Alexander Pawlis. The early stimulus to design this experiment was given by Beata Kardynal and is greatly acknowledged. Further, we thank Fabian Schmidt for early support in the setup of the experiment, and Till Ockenfels for continuous advice in the construction of the high-pressure cell. We acknowledge funding by Deutsche Forschungsgemeinschaft DFG through grant INST 217/978-1 FUGG and through the Cluster of Excellence ML4Q (EXC 2004/1 - 390534769).\\

Data underlying the results presented in this paper are not publicly available at this time but may be obtained from the authors upon reasonable request.
\end{acknowledgments}

\bibliographystyle{apsrev}
\bibliography{bib}

\end{document}